\documentclass[onecolumn,10pt]{article}
\usepackage{amsmath}
\usepackage{amsfonts}
\usepackage{amssymb}
\usepackage{amsthm}
\usepackage{times}
\usepackage{cancel}
\usepackage{dsfont}
\usepackage{graphics}

\begin{document}

\date{}
\title{Foreword}
\author{} 
\maketitle

\emph{Understanding Quantum Raffles} was inspired by \emph{Bananaworld}, as the authors say, but it is very much more than that. My initial aim in writing \emph{Bananaworld} was to de-mystify quantum entanglement for non-physicists---as Schr\"{o}dinger remarked, `\emph{the} characteristic trait of quantum mechanics, the one that enforces its entire departure from classical lines of thought.' I wanted to show that entanglement is essentially a new sort of nonlocal correlation, explain why it is puzzling, and point out how it can be used as a resource. 
The device I used to exhibit entanglement was the Popescu-Rohrlich nonlocal box, or PR-box, which I dramatized as a pair of bananas that each acquires one of two possible tastes when peeled in one of two allowable ways, from the stem end or the top end. The PR-box correlation is a superquantum correlation but can be expressed quite simply, without the mathematical machinery of quantum mechanics. It has all the puzzling features of quantum entanglement and, with a little poetic license, can even be exploited to show how entanglement works to enable feats like quantum teleportation, unconditional security in quantum cryptography, and apparently exponential speed-up in quantum computation.

In spite of the bananas, the book did not turn out to be the sort of thing you could pick up and enjoy over a beer. So I wrote \emph{Totally Random: Why Nobody Understands Quantum Mechanics} with my daughter, Tanya Bub. \emph{Totally Random} deals with some of the topics discussed in \emph{Bananaworld}, but in a way that's much more accessible and, we hoped, fun to read. We presented the book as  `a serious comic on entanglement'---serious because we felt that the general reader could come away with a real understanding of entanglement: what it is, what the patriarchs of quantum mechanics have said about it, and what what you can do with it. The authors of \emph{Understanding Quantum Raffles}---the three Mikes---have evidently also given a great deal of thought to pedagogical issues. While some of the discussion, notably Chapter 4, tackles advanced material, a major part of the book, especially Chapters 2 and 3, is clearly intended for the general reader, so if you want to  understand what is really new and interesting about quantum mechanics, this is the book to read.

In \emph{Bananaworld}, I brought out the difference between classical and quantum mechanics by considering to what extent it is possible to simulate a PR-box correlation with various resources, classical or quantum. Bell's nonlocality proof amounts to a demonstration that two separated agents, Alice and Bob, restricted to classical, and so local resources (effectively what computer scientists call `shared randomness'), can achieve an optimal success rate of no more than $75\%$. If Alice and Bob are allowed to use quantum resources, entangled pairs of photons or electrons, they can do better, about $85\%$. Equipped with PR-boxes, they can, of course, achieve a $100\%$ success rate. Another way to put this is in terms of the Clauser-Horne-Shimony-Holt (CHSH) inequality for two bivalent Alice-observables and two bivalent Bob-observables. The CHSH correlation for the four pairs of observables is constrained to values between $-2$ and $2$ for for local classical correlations, between $-2\sqrt{2}$ and $2\sqrt{2}$ for quantum correlations, and between $-4$ and $4$ for PR-box correlations, which are maximal for correlations that do not allow instantaneous signaling. Geometrically, as Pitowsky showed,\footnote{I. Pitowsky, `On the geometry of quantum correlations,' \emph{Physical Review} A 77, 062109 (2008).} the classical or local correlations for this case can be represented by the points in an $8$-dimensional polytope with facets characterized by the CHSH inequality and similar inequalities, the quantum correlations by the points in a convex set that includes the polytope, and the no-signaling correlations by a polytope that includes the quantum convex set. 

The three Mikes do something brilliantly different. Instead of the CHSH inequality, they consider the Mermin inequality for three bivalent observables for each agent. In terms of bananas, Alice and Bob peel their bananas in one of three possible ways associated with three directions in which they are required to hold their bananas while peeling. This complication, which I blush to admit I first thought was pointless, results in a tetrahedron for the classical or local correlations, an elliptope for the quantum convex set (a `fat' tetrahedron that includes the classical tetrahedron), and a cube for the no-signaling correlations---easily visualizable in three dimensions. The  three Mikes produce two derivations for the non-linear inequality characterizing the elliptope: a derivation `from within' quantum mechanics, which uses the Born rule for probabilities, and a derivation `from without,' which follows work by Yule in the late 19th century on Pearson correlation coefficients. In Yule's derivation, the inequality is a general constraint on correlations between three random variables. In the `proof from without,' the random variables are the eigenvalues of Hilbert space operators representing observables and the constraint follows quite generally, without assuming the Born rule for quantum probabilities. 

The Mermin inequality refers to spin-1/2 particles in the singlet state. Remarkably, it turns out that singlet state quantum correlations are confined to the elliptope even for higher spin values, while the tetrahedron for local classical correlations is replaced by a succession of polyhedra with more and more facets for higher spins, approaching the elliptope in the limit of infinite spin. All this is beautifully illustrated in $3$-dimensional visualizations. The analysis is particularly impressive because it shows clearly and precisely how classical and quantum correlations are related in this particular case. 

This is certainly the first book in which the word `Bubism' appears. The three Mikes use the term to refer to `an interpretation of quantum mechanics along the lines of \emph{Bananaworld}, belonging to the same lineage, or so we will argue, as the much-maligned Copenhagen interpretation.' \emph{Bananaworld} began as a discussion of entanglement, but as I wrote the book it evolved into a way of thinking about the transition from classical to quantum mechanics. The three Mikes have taken this perspective and articulated and developed it into an interpretation that I fully endorse but which owes as much to their careful analysis of the conceptual issues as my own thinking. 

I added the last chapter to \emph{Bananaworld},  `Making Sense of it All,' because I thought I should say something about the measurement problem of quantum mechanics as it is usually understood, and how various  interpretations propose to solve the problem. But the chapter doesn't fit well with the rest of the book, which, taken as a whole, was already an attempt to make sense of it all. The revised version in the paperback edition is an improvement, but not entirely satisfactory.  Chapter 6 of \emph{Understanding Quantum Raffles}, on interpreting quantum mechanics, nails it.

Here, following the account by the three Mikes, is how I now see the view they call Bubism. Quantum mechanics began with Heisenberg's unprecedented move to `reinterpret' classical quantities like position and momentum as noncommutative. In a commutative algebra, the $2$-valued quantities, representing propositions that can be true or false, form a Boolean algebra. A Boolean algebra is isomorphic to a set of subsets of a set, with the Boolean operations corresponding to the union, intersection, and complement of sets. The conceptual significance of Heisenberg's proposal lies in replacing the Boolean algebra of subsets of classical phase space, where the points represent classical states and subsets represent ranges of values of dynamical variables, with a non-Boolean algebra. Later, following the Born-Heisenberg-Jordan \emph{Dreim\"{a}nnerarbeit} and further developments by Dirac, Jordan, and von Neumann, this non-Boolean algebra was formalized as the algebra of closed subspaces of Hilbert space, a vector space over the complex numbers, or equivalently a projective geometry. So the transition from classical to quantum mechanics is, formally, the transition from a Boolean algebra of subsets of a set to a non-Boolean algebra of subspaces of a vector space.

In his 1862 work `On the Theory of Probabilities,' George Boole characterized a Boolean algebra as capturing `the conditions of possible experience.' Classical theories are Boolean theories. The non-Boolean algebra of quantum mechanics (for Hilbert spaces of more than two dimensions) can be pictured as a family of Boolean algebras that are `intertwined,' to use Gleason's term,\footnote{A. N. Gleason, `Measures on the closed subspaces of Hilbert space,' \emph{Journal of Mathematics and Mechanics} 6, 885--893 (1957). The term is used to refer to intertwined orthonormal sets, which are Boolean algebras, on p. 886.} or `pasted together,' in such a way that the whole family can't be embedded into a single Boolean algebra.\footnote{Kochen and Specker proved non-embeddability for the `partial Boolean algebra' of subspaces of a Hilbert space of more than two dimensions in S. Kochen and E.P. Specker, `On the problem of hidden variables in quantum mechanics,' \emph{Journal of Mathematics and Mechanics} 17, 59--87 (1967). Bell proved a related result as a corollary to Gleason's theorem in J.S. Bell, `On the problem of hidden variables in quantum mechanics,' \emph{Reviews of Modern Physics} 38, 447--452 (1966), reprinted in J.S. Bell, \emph{Speakable and Unspeakable in Quantum Mechanics} (Cambridge University Press, Cambridge, 1987).}  So in a quantum theory, the single Boolean algebra of a classical theory is replaced by a family of Boolean algebras, in effect, a family of Boolean perspectives or Boolean frames associated with different incompatible measurement experiences. The upshot, as von Neumann pointed out, is that quantum probabilities are `perfectly new and \emph{sui generis} aspects of physical reality'\footnote{From an unpublished manuscript `Quantum logics (strict- and probability-logics),' reviewed in A.H. Taub in \emph{John von Neumann: Collected Works} (Macmillan, New York, 1962), volume 4, pp. 195--197.} and `uniquely given from the start.' 

The sense in which quantum probabilities are `uniquely given from the start' is explained in an address by von Neumann on `unsolved problems in mathematics'  to an international congress of mathematicians in Amsterdam, September 2--9, 1954.\footnote{In Mikl\'os R\'edei and Michael St\"{o}ltzner (eds.), \emph{John von Neumann and the Foundations of Quantum Mechanics},  pp. 231--246 (Kluwer Academic Publishers, Dordrecht, 2001). The quoted passage is on pp, 244--245. Also quoted (without the last sentence) in M. R\'edei, ```Unsolved Problems in Mathematics' J. von Neumann's address to the International Congress of Mathematicians Amsterdam, September 2--9, 1954,' \emph{The Mathematical Intelligencer} 21, 7--12 (1999).} Here is the relevant passage:
\begin{quotation}
Essentially if a state of a system is given by one vector, the transition probability in another state is the inner product of the two which is the square  of the cosine of the angle between them {[}sic{]}.\footnote{Von Neumann evidently meant to say that the transition probability is the  square of the (absolute value of) the inner product, which is the square  of the cosine of the angle between them.}  In other words, probability corresponds precisely to introducing the angles geometrically. Furthermore, there is only one way to introduce it. The more so because in the quantum mechanical machinery the negation of a statement, so the negation of a statement which is represented by a linear set of vectors, corresponds to the orthogonal complement of this linear space. And therefore, as soon as you have introduced into the projective geometry the ordinary machinery of logics, you must have introduced the concept of orthogonality. This actually is rigorously true and any axiomatic elaboration of the subject bears it out. So in order to have logics you need in this set a projective geometry with a concept of orthogonality in it. 

In order to have probability all you need is a concept of all angles, I mean angles other than $90^{\circ}$. Now it is perfectly quite true that in geometry, as soon as you can define the right angle, you can define all angles. Another way to put it is that if you take the case of an orthogonal space, those mappings of this space on itself, which leave orthogonality intact, leave all angles intact, in other words, in those systems which can be used as models of the logical background for quantum theory, it is true that as soon as all the ordinary concepts of logic are fixed under some isomorphic transformation, all of probability theory is already fixed. 

What I now say is not more profound that saying that the concept of a priori probability in quantum mechanics is uniquely given from the start. 
\end{quotation}

In \emph{Bananaworld}, I defended what I called an `information-theoretic' interpretation of quantum mechanics. The term is perhaps unfortunate. In the first place, it invites objections like those by Bell: `\emph{Whose} information? Information about \emph{what}?'\footnote{J.S. Bell, `Against measurement,' in \emph{Physics World} 8, 33--40 (1990). The comment is on p. 34.} In the second place, the emphasis should be on probability, as the three Mikes make clear, with the understanding that information theory is a branch of probability theory specifically concerned with probabilistic correlations. 

If relativity is about space and time, quantum mechanics is \emph{about probability}, in the sense that quantum probabilities are `\emph{sui generis}' and `uniquely given from the start' as an aspect of the kinematic structure of the theory and are not imposed from outside as a measure of ignorance, as in classical theories, where probability is a measure over phase space. In this new framework, new sorts of nonlocal probabilistic correlations associated with entanglement are possible, which makes quantum information fundamentally different from classical information. In a Boolean theory such correlations are impossible without introducing what Einstein called `spooky' action at a distance.

Quantum probabilities are revealed in measurement, and a measurement is associated with the selection of a particular Boolean frame in the family of Boolean algebras that `captures the conditions of possible experience.'  In terms of observables,  a measurement involves the selection of a basis of commuting observables in Hilbert space.  As a consequence, the observer is no longer `detached,' unlike the observer in classical mechanics, as Pauli observed.\footnote{M. Born, \emph{The Born-Einstein Correspondence} (Walker and Co., London, 1971). Pauli talks about the classical ideal of the `detached observer'  in a letter to Born dated March 30, 1954 on p. 218.} The measurement outcome is a random assignment of truth values to the elements in the Boolean frame, or a random assignment of values to the observables in the corresponding basis. What's puzzling, from a Boolean perspective, is that measurement in a non-Boolean theory is not passive---not just `looking' and registering what's there in a passive sense. Measurement must produce a change in the description, and that's not how we are used to thinking of measurement in a Boolean theory. Here's how Schr\"{o}dinger puts it:\footnote{`Die gegenw\"{a}rtige Situation in der Quantenmechanik,' \emph{Die Naturwissenschaften} 48, 807--812; 49, 823--828, 844--849 (1935). The quotation is from p. 826. The translation is by John Trimmer, \emph{Proceedings of the American Philosophical Society} 124, 323---338 (1980).}
\begin{quote}
(1) The discontinuity of the expectation-catalog {[}the quantum pure state{]} due to measurement is \emph{unavoidable}, for if measurement is to retain any meaning at all then the \emph{measured value}, from a good measurement, \emph{must} obtain. (2) The discontinuous change is certainly \emph{not} governed by the otherwise valid causal law, since it depends on the measured value, which is not predetermined. (3) The change also definitely includes (because of `maximality' {[}the `completeness' of the quantum pure state{]}) some \emph{loss} of knowledge, but knowledge cannot be lost, and so the object \emph{must} change---\emph{both} along with the discontinuous changes and \emph{also}, during these changes, in an unforeseen, \emph{different} way.
\end{quote}

Quantum probabilities don't simply represent ignorance about what is the case. Rather, they represent a new sort of ignorance about something that doesn't yet have a truth value, something that simply isn't one way or the other before we measure, something that requires us to act and do something that we call a measurement before nature supplies a truth value---and removes the truth values of incompatible propositions that don't belong to the same Boolean frame, associated with observables that don't commute with the measured observable. 

Schr\"{o}dinger calls the measurement problem `the most difficult and most interesting point of the theory.'\footnote{\emph{ibid.}, p. 826.} As the three Mikes aptly put it, the measurement problem is \emph{a feature of quantum mechanics} as a non-Boolean theory, \emph{not a bug}.

Interpretations of quantum mechanics that oppose the Copenhagen interpretation begin with  Schr\"{o}dinger's wave theory as conceptually fundamental, rather than Heisenberg's algebraic formulation of quantum mechanics, and propose dynamical solutions to what then seems to be a problem: how does what we do when we perform a measurement by manipulating some hardware in a laboratory  select a Boolean frame in Hilbert space, a basis of observables that have definite values, and what explains the particular assignment of truth values to the elements in the Boolean frame, or the particular assignment of values to observables.

Bohm's theory tells a one-world Boolean story: position in configuration space is always definite, associated with a Boolean algebra, and other quantities become definite through correlation with position via the measurement dynamics. The problem here, as Bell showed, is that Bohm's theory is nonlocal in configuration space, allowing instantaneous action at a distance, which Einstein regarded as `spooky'\footnote{M. Born, \emph{op. cit.}.  The term is used in a letter from Einstein to Born dated March 3, 1947 on p. 158.}and so non-physical (although averaging over the Born distribution hides the nonlocality). I suspect that it was for this reason that Einstein dismissed Bohm's theory as `too cheap for me' in a letter to Born.\footnote{M. Born, \emph{op. cit.}  The comment is on p. 192 in a letter from Einstein to Born dated May 12, 1952.}

The Everett interpretation tells a multi-world Boolean story in which everything that can happen does happen in some Boolean world. This avoids having to explain why \emph{this} measurement outcome rather than \emph{that} measurement outcome, since every possible outcome actually occurs in some world. The trick is to show how this fits Schr\"{o}dinger's wave theory of quantum mechanics.     There is no spooky action at a distance in the Everettian interpretation, but the measurement problem appears as the basis problem: how to explain the selection of a particular basis with respect to which the multiplicity associated with `splitting into many worlds' occurs in a measurement process. Everettians solve the basis problem by appealing to the dynamics of environmental decoherence: as the environment becomes increasingly entangled with the measuring apparatus, it  becomes more and more difficult, but not in principle impossible, to distinguish an entangled state from the corresponding mixture with respect to a particular coarse-grained basis. Quantum probabilities with respect to the elements of this basis are explained in terms of the decision theory of an agent-in-a-world about to make a measurement.  Even granting decoherence as an effective solution to the basis problem, it seems contrived to interpret  the `perfectly new and \emph{sui generis} aspects of physical reality,' the Hilbert space probabilities that are `uniquely given from the start,' in this way.

\emph{Understanding Quantum Raffles} is likely to be a classic in the foundational literature on quantum mechanics. The three Mikes have produced  an exceptionally lucid book on quantum foundations that is also suitable for readers, with some tolerance for basic algebra and geometry, who are looking for answers to conceptual questions that are   typically glossed over in standard courses on quantum mechanics. 

\bigskip\bigskip 
\noindent Jeffrey Bub\\
Philosophy Department\\
Institute for Physical Science and Technology\\	
Joint Center for Quantum Information and Computer Science	\\	
University of Maryland, College Park

\end{document}